\documentclass[aps, prl, twocolumn, floatfix, superscriptaddress]{revtex4-1} 
\usepackage{graphicx}
\usepackage{epstopdf}
\usepackage{bm, amsmath, amssymb}

\begin{document}

\title{Non-Poissonian Quantum Jumps of a Fluxonium Qubit due to Quasiparticle Excitations}

\author{U. Vool}
\email{uri.vool@yale.edu}
\affiliation{Department of Applied Physics and Physics, Yale University, New Haven, CT 06520}
\author{I. M. Pop}
\affiliation{Department of Applied Physics and Physics, Yale University, New Haven, CT 06520}
\author{K. Sliwa}
\affiliation{Department of Applied Physics and Physics, Yale University, New Haven, CT 06520}
\author{B. Abdo}
\altaffiliation{Current address: IBM T. J. Watson Research Center, Yorktown Heights, New York 10598, USA.} 
\affiliation{Department of Applied Physics and Physics, Yale University, New Haven, CT 06520}
\author{C. Wang}
\affiliation{Department of Applied Physics and Physics, Yale University, New Haven, CT 06520}
\author{T. Brecht}
\affiliation{Department of Applied Physics and Physics, Yale University, New Haven, CT 06520}
\author{Y. Y. Gao}
\affiliation{Department of Applied Physics and Physics, Yale University, New Haven, CT 06520}
\author{S. Shankar}
\affiliation{Department of Applied Physics and Physics, Yale University, New Haven, CT 06520}
\author{M. Hatridge}
\affiliation{Department of Applied Physics and Physics, Yale University, New Haven, CT 06520}
\author{G. Catelani}
\affiliation{Peter Gr{\"{u}}nberg Institut (PGI-2), Forschungszentrum J{\"{u}}lich, 52425 J{\"{u}}lich, Germany}
\author{M. Mirrahimi}
\affiliation{Department of Applied Physics and Physics, Yale University, New Haven, CT 06520}
\affiliation{ INRIA Paris-Rocquencourt, Domaine de Voluceau, BP105, 78153 Le Chesnay cedex, France}
\author{L. Frunzio}
\affiliation{Department of Applied Physics and Physics, Yale University, New Haven, CT 06520}
\author{R. J. Schoelkopf}
\affiliation{Department of Applied Physics and Physics, Yale University, New Haven, CT 06520}
\author{L. I. Glazman}
\affiliation{Department of Applied Physics and Physics, Yale University, New Haven, CT 06520}
\author{M. H. Devoret}
\affiliation{Department of Applied Physics and Physics, Yale University, New Haven, CT 06520}

\date{\today}

\begin{abstract}

As the energy relaxation time of superconducting qubits steadily improves, non-equilibrium quasiparticle excitations above the superconducting gap emerge as an increasingly relevant limit for qubit coherence. 
We measure fluctuations in the number of quasiparticle excitations by continuously monitoring the spontaneous quantum jumps between the states of a fluxonium qubit, in conditions where relaxation is dominated by quasiparticle loss. Resolution on the scale of a single quasiparticle is obtained by performing quantum non-demolition projective measurements within a time interval much shorter than $T_1$, using a quantum limited amplifier (Josephson Parametric Converter). The quantum jumps statistics switches between the expected Poisson distribution and a non-Poissonian one, indicating large relative fluctuations in the quasiparticle population, on time scales varying from seconds to hours. This dynamics can be modified controllably by injecting quasiparticles or by seeding quasiparticle-trapping vortices by cooling down in magnetic field. 

\end{abstract}

\maketitle

A mesoscopic superconducting circuit, of typical size smaller than $1\: \mathrm{mm^3}$, cooled to a temperature well below the superconducting gap should be completely free of thermal quasiparticle (QP) excitations.
However, in the last decade there has been growing experimental evidence that the QP density at low temperatures saturates to values orders of magnitude above the value expected at thermal equilibrium\cite{Aumentado2004,Ferguson2006,Martinis2009,Shaw2008,Visser2011}.
These non-equilibrium QP excitations limit the performance of a variety of superconducting devices, such as single-electron turnstiles\cite{Pekola2008}, kinetic inductance\cite{Day2003,Monfardini2012} and quantum capacitance\cite{Stone2012} detectors, micro-coolers\cite{Giazotto2006,Rajauria2009}, as well as Andreev bound state nano-systems\cite{Bretheau2013a,Levenson-Falk2014}.
Moreover, QP's are an important intrinsic decoherence mechanism for superconducting two level systems (qubits)\cite{Lutchyn2005,Lenander2011,Catelani2011a,Sun2012,Wenner2013,Riste2013}. 
In particular, a recent experiment performed on the fluxonium qubit showed energy relaxation times in excess of $1\: \mathrm{ms}$, limited by QP's\cite{Pop2014}. 
Surprisingly, the sources generating these QP excitations are not yet positively identified.
The measurement of non-equilibrium QP dynamics at low temperatures could provide insight into their origin as well as an efficient tool to quantify QP suppression solutions.

In this letter, we show that the quantum jumps\cite{Vijay2011} of a qubit whose lifetime is limited by QP tunneling, such as the fluxonium artificial atom, can serve as a sensitive probe of QP dynamics. 
A jump in the state of the qubit indicates an interaction of the qubit with a QP, and therefore fluctuations in the rate of quantum jumps are directly linked to changes in QP number. 
Tracking the state of the qubit in real time requires fast, single-shot projective measurement with minimal added noise, made possible by the advent of quantum-limited amplifiers\cite{Castellanos-Beltran2008,Bergeal2010,Hatridge2011}.
In this work, we use a Josephson Parametric Converter (JPC) quantum limited amplifier\cite{Bergeal2010,Abdo2013} to monitor the state of our qubit with a resolution of 5 $\mu$s, two orders of magnitude faster than the qubit lifetime.
We find that the qubit jump statistics fluctuates between Poissonian and non-Poissonian, corresponding to a change in the QP number. Surprisingly, these fluctuations do not average over timescales ranging from seconds to hours.
The quantum jumps we measure in this work are driven by a few QP's in the entire device at any given time. In a related work, the dynamics of a population of a few thousands of QP's is probed by $T_1$ measurements of a transmon qubit\cite{Wang2014}. 

The fluxonium qubit\cite{Manucharyan2009} (Fig. 1a) consists of a Josephson junction shunted by a superinductor\cite{Manucharyan2011,Brooks2013}, which is itself an array of large Josephson junctions\cite{Masluk2012}. 
An optical image of the fluxonium sample coupled to its readout antenna is shown in Fig.~1b.
An applied external flux $\Phi_{\mathrm{ext}}$ strongly affects the fluxonium spectrum, energy eigenstates, and its susceptibility to different loss mechanisms.
The overall quality factor $Q$ of the fluxonium is given by: 
\begin{eqnarray}
\frac{1}{Q} = \sum\limits_{x} \eta_x \frac{p_x}{Q_x} \label{eq:transitioneff}
\end{eqnarray}
where $Q_x$ is the quality factor of the material involved in loss mechanism $x$, $p_x$ is its participation ratio and $\eta_x$ is the oscillator strength of the qubit transition induced by $x$.              
Fig. 1c shows $\eta_x$ as a function of external flux for three main loss mechanisms - capacitive, inductive and QP tunneling across the small junction. 
The main inductive loss mechanism for the fluxonium is due to QP tunneling across the array junctions.
Note that around $\Phi_{\mathrm{ext}}=\Phi_0/2$ the fluxonium qubit becomes insensitive to loss due to QP tunneling across the small junction and maximally sensitive to loss due to QP tunneling across the array junctions.

\begin{figure*} [htp]

 \includegraphics[angle = 0, width = \textwidth]{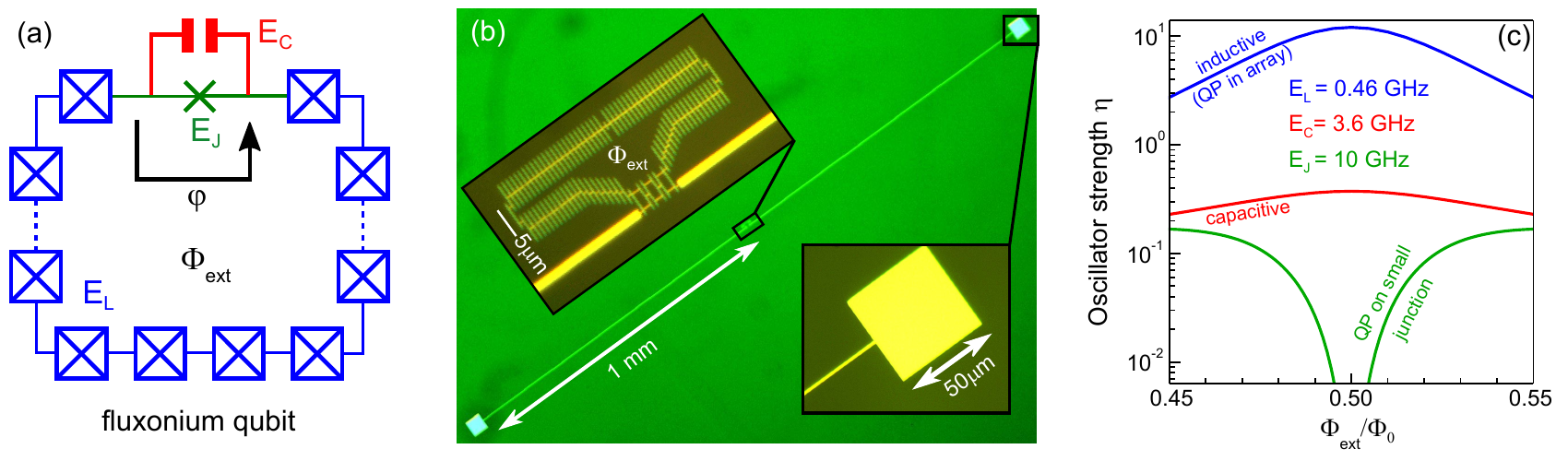}
 \caption{\label{fig1}  The fluxonium qubit. (a) Electrical circuit schematic. The small junction, which is modeled by an ideal tunnel junction (green) in parallel with a capacitor (red), is shunted by an array of large Josephson junctions (blue). 
 (b) Optical microscope image of the fluxonium qubit inductively coupled to the antenna. The top and bottom insets show magnified images of the fluxonium loop and antenna pads, respectively.
The small junction is shunted by a superinductance composed of an array of 95 Josephson junctions, enclosing a magnetic flux $\Phi_{\mathrm{ext}}$. 
 (c) The oscillator strength $\eta$ (see Eq. \ref{eq:transitioneff}) of different qubit decay mechanisms vs. the applied external flux $\Phi_{\mathrm{ext}}$. The corresponding capacitive, inductive and Josephson energies, defined for the fluxonium artificial atom in Ref. \cite{Manucharyan2009}, are shown as $E_C$, $E_L$ and $E_J$ respectively.
 }
 \end{figure*}
The insensitivity of the fluxonium qubit to QP tunneling across the small junction was 
demonstrated by the measurement of a sharp $T_1$ increase to values above $1\:\mathrm{ms}$ in the vicinity of $\Phi_0/2$\cite{Pop2014}.
In addition, non-exponential decay curves were occasionally measured, suggesting
a fluctuating QP population.
To gain access to these fluctuations, we improved the readout setup used in Ref. \cite{Pop2014} by adding a JPC amplifier, thus increasing the signal-to-noise ratio of the setup by a factor of 10. 
A schematic of the measurement setup is presented in Fig.~2a.

To monitor the state of the qubit, we apply a continuous wave drive at the cavity resonance corresponding to an average photon population $\bar{n} = 2.5$. This value is a compromise between fast measurement
and the effect of cavity photons which reduce the qubit lifetime and saturate the JPC output\cite{Supplementary}.  
In Fig. 2b we show a histogram of measured $I,Q$ quadratures at flux bias point $\Phi_{\mathrm{ext}}=\Phi_0/2$ where the qubit frequency is $\omega_{ge}/2\pi = 665\: \mathrm{MHz}$.
The measured distributions corresponding to the ground/excited states of the fluxonium qubit (right/left) are separated by 5 standard deviations $\sigma$. 
The relative population of the fluxonium in its excited state (33\%) corresponds to an effective temperature of 45~mK.

A few examples of measured qubit quantum jump traces are shown in Fig.~2c. 
To estimate the state of the qubit (orange) from the time trace of quadrature $I$ (blue) we apply a two-point filter. The filter declares a jump in the qubit state
if the quadrature value crosses a threshold set $\sigma/2$ away from the jump destination. Otherwise the qubit is declared to remain in its previous state.
The traces suggest there are two regimes with distinctly different jump statistics. There appear to be ``quiet'' times with few jumps and long intervals between them (on the order of 1 ms) and ``noisy'' times
with many rapid consecutive jumps (less than 100 $\mu$s apart).

\begin{figure} [htp]

\includegraphics[angle = 0, width = \columnwidth]{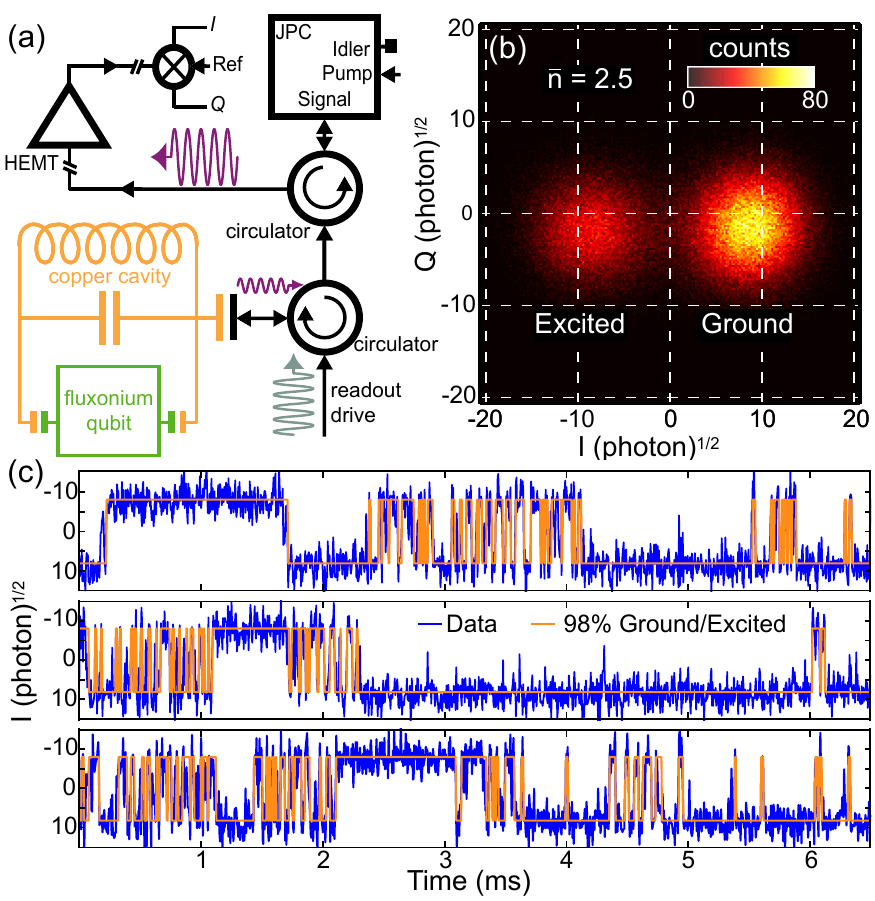}
\caption{\label{fig2}   (a) Circuit diagram of the measurement setup. The fluxonium qubit (green) is dispersively coupled to the readout cavity (orange) through the antenna (see Fig.~1b). 
The readout signal reflected from the cavity is pre-amplified using a JPC, then it is routed through a commercial HEMT amplifier at $4 \:\mathrm{K}$ and demodulated using a heterodyne interferometry setup at room temperature.
(b) Histogram of measured $I,Q$ quadratures for the fluxonium in equilibrium with its environment at $\Phi_{\mathrm{ext}}=\Phi_0/2$ ($\omega_{ge}/2\pi=665\:\mathrm{MHz}$).
Each count corresponds to $5\: \mathrm{\mu s}$ of integration and the total number of counts is 200,000.
The two distinct peaks in the $I,Q$ plane correspond to the ground (right) and 1st excited (left) states of the qubit. Their relative amplitudes give an effective temperature of $45$~$\mathrm{mK}$.
(c) Three examples of measured quantum jump traces corresponding to the time evolution of the $I$ quadrature for the same measurement presented in (b). 
We show the raw traces in blue and an estimate of the fluxonium qubit state calculated using a two-point filter (see text) in orange. 
Note that the characteristic time between jumps is not constant throughout the record.
}
\end{figure}
Uncorrelated quantum jumps obey Poisson statistics, leading to an exponential distribution of the time spent in the ground or excited state $p_\textnormal{\tiny \textsc{P}}(\tau) = \frac{1}{\bar{\tau}}e^{-\tau / \bar{\tau}}$ 
where $\bar{\tau}$ is the mean time spent in the ground or excited state. 
To enhance the visibility of deviations from Poisson statistics, which would merely show up as non-exponential decrease of $p(\tau)$, we depict the distribution $\tau p(\tau)$ instead.
In Fig. 3a and 3b we show two different second-long measurements of $\tau p(\tau)$ distributions for the ground (blue) and excited (red) states, histogrammed with logarithmic bins.
The dashed lines correspond to the distribution predicted by Poisson statistics
with $\bar{\tau}$ taken as the measured average time either in the ground (blue) or excited (red) state (see supplementary material\cite{Supplementary} for a detailed definition).
There is significant deviation between the two measurements. In Fig. 3b we show a measurement record which we call ``quiet'', apparently agreeing with Poisson statistics.
The ``noisy'' record in Fig. 3a deviates significantly from the Poisson prediction, with long and short times
appearing considerably more frequently than expected. 
 
In Fig.~3c, the mean time spent in the ground (blue) and excited (red) state is shown as a function of time, over several minutes. Each point corresponds to a 1 second temporal average.
To quantify the deviation of each measurement from Poisson statistics, we calculate the fidelity of the measured histogram to the Poisson prediction $F=\frac{\sum\limits_i\sqrt{M_i P_i}}{\sum\limits_i{M_i}}$, 
where $M_i$ is the measured ground state histogram value of bin $i$ and $P_i$ is the predicted value of bin $i$ for a Poisson process. In Fig.~3d, we plot the deviation from Poisson statistics, $1-F$, corresponding to the measurements in Fig.~3c.
These two figures indicate a correlation between long fluxonium energy lifetimes and agreement with Poisson statistics\cite{Supplementary}. The ``noisy'' seconds appear to have an abundance of short quantum jumps
which distort the Poisson statistics, typical for the ``quiet'' seconds.
Fig. 3e shows $\bar{\sigma}_z$, the mean polarization of the fluxonium qubit for the same measurements. The fluctuations in polarization are not correlated with the 
fluctuations between ``quiet'' and ``noisy'' seconds. The examples in Fig. 3a and 3b were taken for measurements with the same polarization corresponding to a temperature of $49 \: \mathrm{mK}$ 
(highlighted in gray in Fig.~3c,d,e).

\begin{figure} [htp]

\includegraphics[angle = 0, width = \columnwidth]{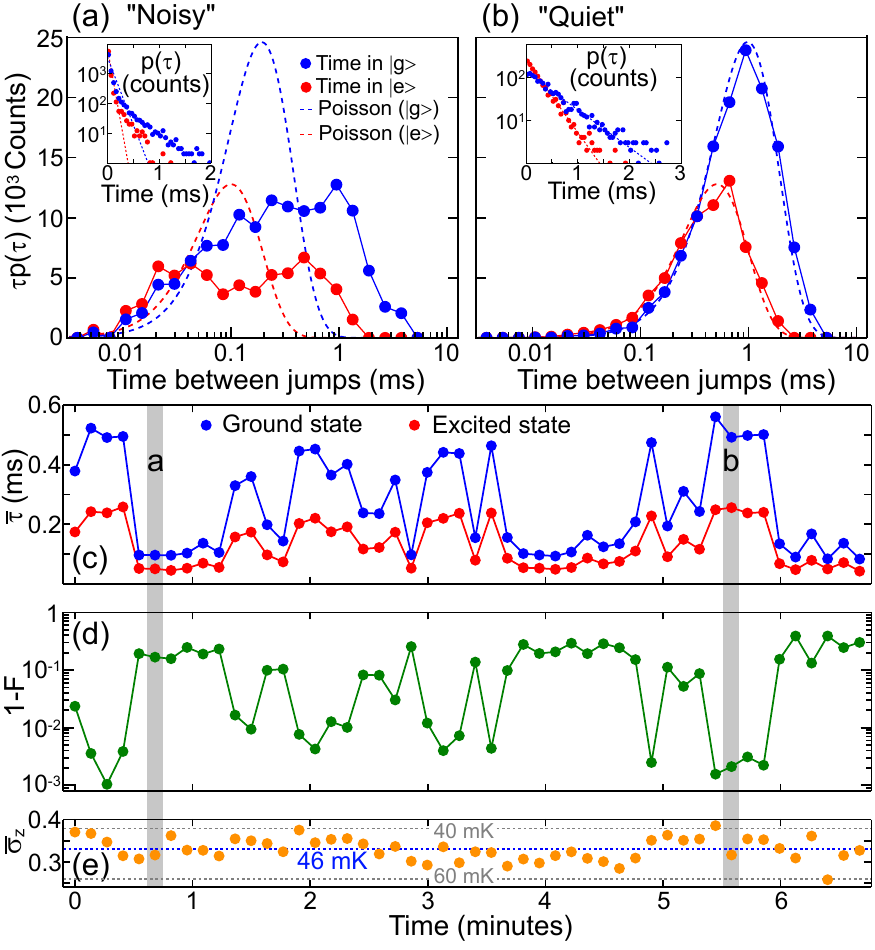}
\caption{\label{fig3}  Measurement of ``quiet" and ``noisy" behavior of the fluxonium quantum jumps.
(a,b) Histograms with logarithmic binning of the time intervals between jumps $\tau$ for the qubit in the ground/excited (blue/red) states, scaled by $\tau$. 
Each count represents a $5\: \mu \mathrm{s}$ time interval.
In dashed lines we plot the Poissonian prediction with the measured mean time interval $\bar{\tau}$.
Each histogram was taken from a 1 s long measurement record.
The insets show the corresponding linear binning histograms proportional to $p(\tau)$.
Data in (b) agree with the Poisson prediction, while in (a) significantly deviate from it.
(c) Measurement of the average time spent by the qubit in the ground (blue) and excited (red) states vs. time. There are significant fluctuations in these values over the course of minutes.
(d) $1-F$ (see text) vs. time, quantifying the deviation between the measured histogram and the corresponding Poisson prediction for the ground state. 
(e) Average polarization of the fluxonium qubit vs. time. The dashed blue line marks the average polarization which corresponds
to a temperature of $46\: \mathrm{mK}$ and the gray dashed lines are markers for $40\: \mathrm{mK}$ and $60\: \mathrm{mK}$. 
Note that the qubit temperature is not correlated with the fluctuations between the ``quiet'' and ``noisy'' intervals.
}
\end{figure}

The susceptibility of the fluxonium qubit at $\Phi_{\mathrm{ext}}=\Phi_0/2$ to loss due to QP in the array suggests that fluctuations in the mean time between qubit jumps 
and their statistics result from the changing QP population. 
To test this hypothesis, we compare our measurements of spontaneous quantum jump traces to measurements in which we modify the number of QP's.
We do this in two ways: generating QP's by applying strong microwave pulses and trapping QP's by cooling in magnetic field.

\begin{figure} [htp]

\includegraphics[angle = 0, width = \columnwidth]{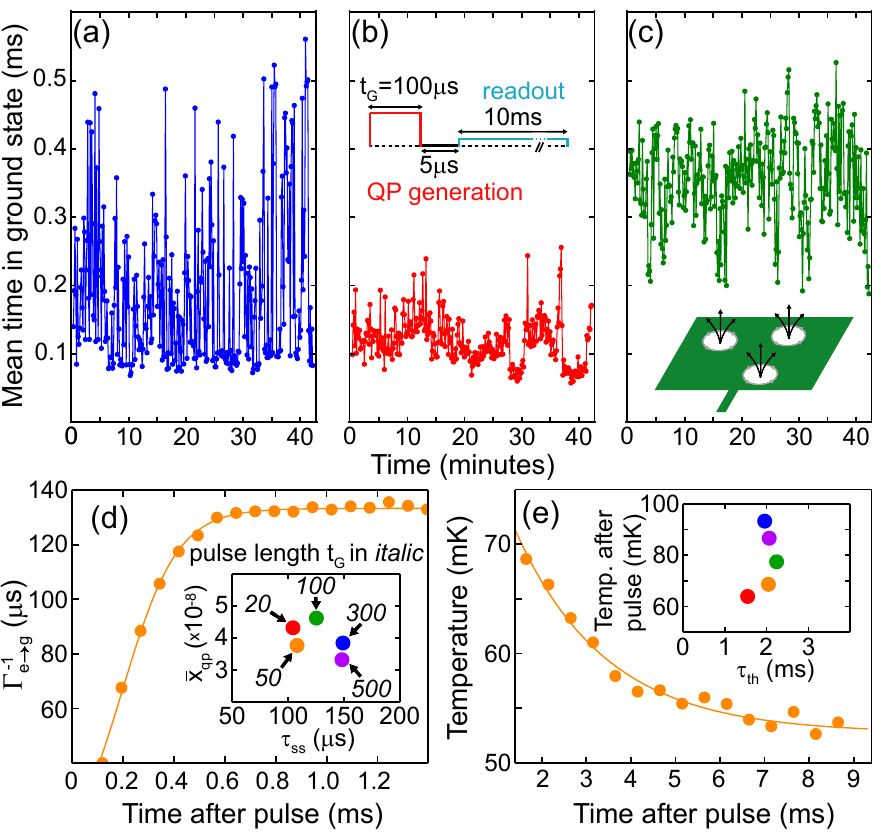}
\caption{\label{fig4}  (a) Mean time in the ground state vs. time. Each point represents an average over 1 second. Similar to the data in Fig. 3c, we observe fluctuations between ``quiet'' and ``noisy'' intervals. 
(b) Mean time in the ground state vs. time in the presence of QP generation pulses. The inset shows the pulse sequence we use. 
A  QP generation pulse of length $t_\textnormal{\tiny \textsc{G}}=100\:\mathrm{\mu s}$ (red) applied at the cavity frequency switches the antenna junctions
into the dissipative regime and generates QP in the junction array. This pulse is followed by $5\:\mathrm{\mu s}$ of wait time (black) and a $10\: \mathrm{ms}$ readout pulse (blue) 
before another QP generation pulse is applied.
(c) Mean time in the ground state vs. time, after cooling in magnetic field. (see text).
(d) Mean time until a jump from the excited to the ground state vs. time after QP generation pulse of length $t_\textnormal{\tiny \textsc{G}}=50\:\mathrm{\mu s}$. The solid line is a fit to exponential equilibration of qubit lifetime (see text). The inset shows the saturation relative QP density $\bar{x}_{qp}$ vs. the time to reach steady-state $\tau_{ss}$ for different QP generation pulse lengths.
(e) Qubit effective temperature vs. time after QP generation pulse of length $t_\textnormal{\tiny \textsc{G}}=50\:\mathrm{\mu s}$. The solid line is a fit to an exponential. Inset shows the temperature after a QP generation pulse vs. the thermal equilibration time $\tau_{th}$ for QP generation pulse lengths corresponding to Fig.~4d.
}
\end{figure}

We created a transient QP population in the array by applying a microwave pulse resonant with the cavity frequency of duration $t_\textnormal{\tiny \textsc{G}}=100\: \mathrm{\mu s}$ and amplitude of order $1$ mV across the antenna, similarly to Ref. \cite{Wang2014}.
After a $5 \: \mathrm{\mu s}$ wait for the
cavity photons to leak out, we monitor quantum jumps for $10$~$\mathrm{ms}$, after which we repeat the cycle.
We estimate that at least $10^{6}$ QP's are generated during each pulse.
In Fig.~4a and b we show a comparison between measurements of the mean time spent in the ground state without and with QP generation pulses.
In the presence of QP generation pulses, the ``quiet'' seconds (higher mean time in the ground state) are suppressed\cite{Supplementary}. 

We reduced the number of QP's by cooling down our sample in a constant magnetic field corresponding to $\Phi_0/2$ in the fluxonium loop. 
Under these conditions, the antenna pads (see Fig.~1b)
are threaded by flux corresponding to several $\Phi_0$. During the field cooldown process the pads can trap vortices\cite{Bardeen1965,Stan2004}, which may act as QP traps due 
to the reduced superconducting gap in their cores\cite{Ullom1998,Peltonen2011,Nsanzineza2014,Wang2014}.
Fig. 4c shows measurements of the mean time spent in the ground state taken after the sample was cooled in magnetic field. 
We observe an increase in the number of ``quiet'' seconds, indicating a reduction in the number of QP's\cite{Supplementary}.
The fluxonium effective temperature changes by less than $10 \:\mathrm{mK}$ between different cooldowns.

Taking advantage of the real time measurement of the qubit relaxation, we can monitor the time evolution of $\bar{\tau}$ after a QP generation pulse.
In Fig. 4d we show the average time spent in the excited state before a jump to the ground state, as a function of time after the QP generation pulse for pulse length $t_ \textnormal{\tiny \textsc{G}} = 50\:\mathrm{\mu s}$.
This yields the equilibration of qubit lifetime as injected QP's leave the junction array. The qubit lifetime eventually saturates to a steady state dominated either by non-thermal QP's or other loss mechanisms. 
The rate to jump from the excited to the ground state  at time $t$ after the pulse is related to the relative QP density by\cite{Catelani2011}:
\begin{eqnarray}
 \Gamma_{e \rightarrow g}(t)= x_{qp}(t) \sqrt{\frac{2\Delta}{\hbar \omega_{ge}}}2\pi \frac{E_{L}}{\hbar} \label{eq:DecayRate}
\end{eqnarray}
where $x_{qp}$ is the ratio of QP's to Cooper pairs in the junction array and $\Delta$ is the superconducting gap.
We fit the lifetime measurements to an exponential model $x_{qp}(t)=\bar{x}_{qp}+(x_{qp}(0)-\bar{x}_{qp})e^{-t/\tau_{ss}}$ from which we extract the time to reach QP steady-state $\tau_{ss} = 125\pm25 \:\mathrm{ \mu s}$ and a non-thermal background QP density $\bar{x}_{qp}=4\pm1\times 10^{-8}$, corresponding to 1-2 QP's in the whole array. Note that the non-Poissonian jump statistics corresponding to the ``noisy'' seconds (see Fig. 3a) show fluctuations in the QP number on the order of their average value, also suggesting the presence of only a few QP's in the whole array.  
This value for $\bar{x}_{qp}$ is an order of magnitude lower than what was measured for the small junction in Ref.~\cite{Pop2014}. The origin of the difference is presently not understood, although one could speculate that QP's in the array more easily diffuse into the antenna.
Note that the value for $\bar{x}_{qp}$ is neither correlated with the QP generation pulse length $t_ \textnormal{\tiny \textsc{G}}$ nor the time to reach steady-state $\tau_{ss}$ (see inset of Fig. 4d). The extracted $\bar{x}_{qp}$ should be treated as an upper bound, since contributions from other decay sources could be present.
Due to the limited dynamic range of our qubit lifetime measurement, of only a factor of 4, we cannot distinguish between different QP removal mechanisms such as trapping, diffusion or recombination\cite{Supplementary}. 
The discrimination between these mechanisms was recently demonstrated in a transmon qubit\cite{Wang2014}.

From the quantum jump traces following a QP generation pulse, we can also extract the average polarization of the qubit and hence its effective temperature.
In Fig. 4e we show the extracted temperature vs. time, starting from $1.4 \: \mathrm{ms}$ after a QP generation pulse, when the QP population has already saturated.
The initial increase in temperature following the QP generation pulse is proportional to the pulse length $t_ \textnormal{\tiny \textsc{G}}$, and it is consistent with an estimated dissipated power of $10^{-10}\:\mathrm{W}$  absorbed in the volume of the sapphire substrate. 
The temperature equilibration time $\tau_{th}$ of several ms is much slower than the sapphire thermalization time and is likely limited by the sapphire-copper contact\cite{Supplementary}.

In conclusion, the distribution of spontaneous quantum jumps of a fluxonium qubit indicates large relative fluctuations in the energy lifetime of this artificial atom. Corresponding changes of the QP density in the superinductor appear to be the natural explanation. This is supported in particular by the increased fluxonium energy lifetime in the presence of QP trapping vortices, which also render the jump statistics Poissonian.
The density of QP's extracted from the measurement does not appear to self-average over periods of seconds, minutes and even hours. This
suggests they originate from sources external to the sample, such as stray infra-red\cite{Martinis2009,Barends2011} or higher energy radiation\cite{Swenson2010,Visser2011}.
In addition, the fluxonium quantum jump statistics resolves a single QP on a $\mu$s timescale, which could be a useful property for a low flux, low energy, particle counting detector.

We acknowledge fruitful discussions with K. Geerlings and S. M. Girvin.
Facilities use was supported by YINQE and NSF MRSEC DMR 1119826. This research was supported by IARPA under Grant No. W911NF-09-1-0369, ARO under Grants No. W911NF-09-1-0514 and W911NF-14-1-0011, NSF under Grants No. DMR-1006060 and DMR-0653377, DOE Contract No. DE-FG02-08ER46482 (LG), and the EU under REA grant agreement CIG-618258 (GC).

\bibliographystyle{apsrev}
\bibliography{QPjumpsbiblio}

\end{document}


\title{Supplementary material for ``Non-Poissonian Quantum Jumps of a Fluxonium Qubit due to Quasiparticle Excitations''}

\author{U. Vool}
\affiliation{Department of Applied Physics and Physics, Yale University, New Haven, CT 06520}
\author{I. M. Pop}
\affiliation{Department of Applied Physics and Physics, Yale University, New Haven, CT 06520}
\author{K. Sliwa}
\affiliation{Department of Applied Physics and Physics, Yale University, New Haven, CT 06520}
\author{B. Abdo}
\affiliation{Department of Applied Physics and Physics, Yale University, New Haven, CT 06520}
\author{C. Wang}
\affiliation{Department of Applied Physics and Physics, Yale University, New Haven, CT 06520}
\author{T. Brecht}
\affiliation{Department of Applied Physics and Physics, Yale University, New Haven, CT 06520}
\author{Y. Y. Gao}
\affiliation{Department of Applied Physics and Physics, Yale University, New Haven, CT 06520}
\author{S. Shankar}
\affiliation{Department of Applied Physics and Physics, Yale University, New Haven, CT 06520}
\author{M. Hatridge}
\affiliation{Department of Applied Physics and Physics, Yale University, New Haven, CT 06520}
\author{G. Catelani}
\affiliation{Peter Grünberg Institut (PGI-2), Forschungszentrum Jülich, 52425 Jülich, Germany}
\author{M. Mirrahimi}
\affiliation{Department of Applied Physics and Physics, Yale University, New Haven, CT 06520}
\affiliation{ INRIA Paris-Rocquencourt, Domaine de Voluceau, BP105, 78153 Le Chesnay cedex, France}
\author{L. Frunzio}
\affiliation{Department of Applied Physics and Physics, Yale University, New Haven, CT 06520}
\author{R. J. Schoelkopf}
\affiliation{Department of Applied Physics and Physics, Yale University, New Haven, CT 06520}
\author{L. I. Glazman}
\affiliation{Department of Applied Physics and Physics, Yale University, New Haven, CT 06520}
\author{M. H. Devoret}
\affiliation{Department of Applied Physics and Physics, Yale University, New Haven, CT 06520}

\date{\today}

\maketitle

\section{Length scales of the fluxonium qubit}
Supplementary Fig. 1 shows detailed Scanning Electron Microscope (SEM) images of the fluxonium qubit used in our experiment, with corresponding length scales. The junctions were fabricated using bridge-free fabrication\cite{Lecocq2011}. The thicknesses of the two aluminum layers are 20 and 30 nm.

 \begin{figure*} [h]

 \includegraphics[angle = 0, width = .8\textwidth]{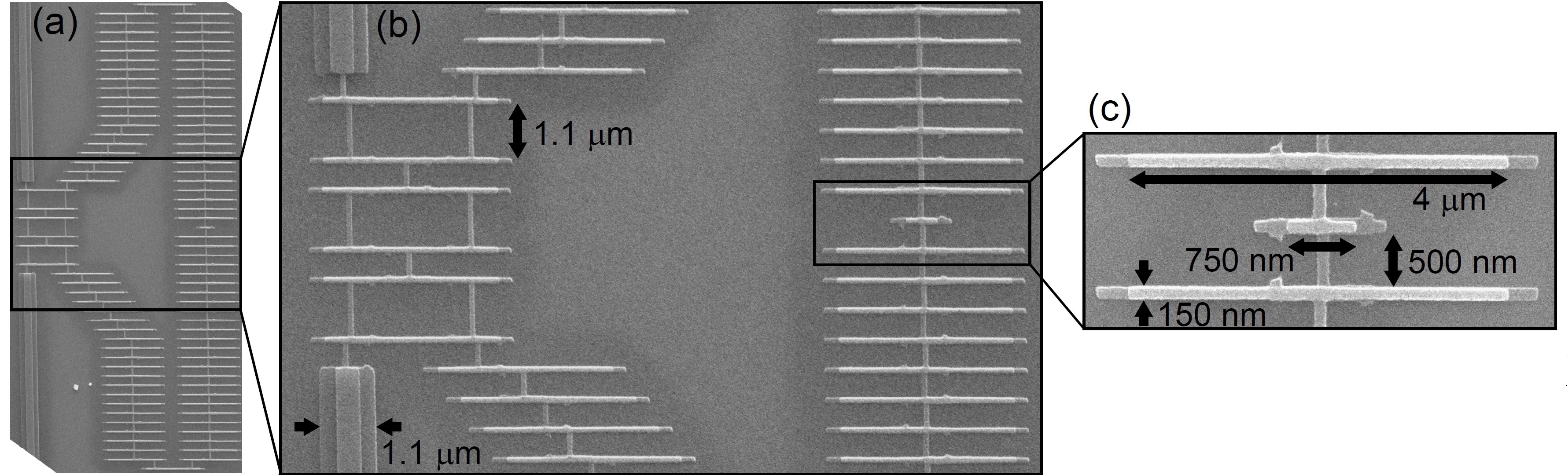}
 \caption{\label{sup_fig1}  
(a) SEM image of the fluxonium loop used in our experiment. The leads extending to the top and bottom of the image connect to the antenna as shown in main text Fig. 2b.
(b) Close-up image showing the SQUID junctions shared between the fluxonium loop and the antenna, big junctions corresponding to the junction array and the small junction. Relevant length scales are shown.
(c) Close-up image showing the small junction and two large junctions with relevant length scales.
 }
 \end{figure*}

\section{Spectroscopy of the fluxonium qubit}
Supplementary Fig. 2 shows a two-tone specroscopy of the fluxonium qubit vs. applied external flux $\Phi_{\mathrm{ext}}$. The qubit frequency $\omega_{ge}/2\pi$ is shown for different external fluxes around the working point $\Phi_{\mathrm{ext}}=\Phi_0/2$. There are no visible avoided level crossings, excluding the presence of spurious two-level systems strongly coupled to the qubit.

 \begin{figure*} [h]

 \includegraphics[angle = 0, width = 0.5\textwidth]{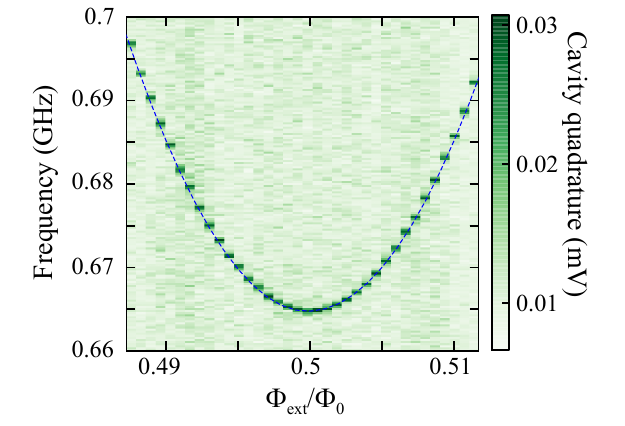}
 \caption{\label{sup_fig2}  
Two-tone spectroscopy of the fluxonium qubit vs. $\Phi_{\mathrm{ext}}$. The fluxonium is probed with a $100 \mathrm{\mu s}$ long saturation pulse at different frequencies, while a measurement tone at the cavity frequency measures the response of the cavity to a change in the fluxonium state. Thus, we measure the fluxonium frequency for different choices of external flux around the working point $\Phi_{\mathrm{ext}}=\Phi_0/2$. The dashed blue line shows the theoretical prediction for the fluxonium frequency vs. external flux, calculated from the values in main text Fig. 1c.
 }
 \end{figure*}

\section{Choice of $\bar{n}$ and measurement time}
The continuous wave drive on the readout cavity affects the system in several ways.
A weak drive will lead to small separation in phase space between the cavity coherent states corresponding to the ground and excited state of the qubit. The separation can be improved by inceasing the integration time, with the cost of reduced time resolution.
On the other hand, an overly strong drive is also problematic as it can saturate the JPC and thus reduce its gain. It also reduces the lifetime of the qubit, an effect that is only qualitatively understood\cite{Boissonneault2008,Ginossar2014,Slichter2012}.
For the fluxonium qubit measured in this work, the lifetime is reduced by 25\% for $\bar{n} = 2.5$ photons in the readout cavity, the value chosen for the data shown in the main text.
Note that although the qubit lifetime decreases due to the presence of photons in the cavity, the quantum jump dynamics is still dominated by QP dissipation. Changes in the QP density induce fluctuations of a factor of 5 in the lifetime of the qubit.

We can relate the separation between the cavity states,  to different system parameters using the relation\cite{Hatridge2013}:
\begin{eqnarray}
\frac{I}{\sigma}=\sqrt{2\bar{n}\kappa T_m \eta}\frac{\chi}{\sqrt{\chi^2+\kappa^2}} \label{eq:Measparams}
\end{eqnarray}
where $I$ is half the difference between the cavity states (see main text Fig. 2b), $\sigma$ is the standard deviation of each state, $\bar{n}$ is the average number of photons in the readout cavity, $\kappa$ is the cavity linewidth and $T_m$ is the measurement integration time.
The qubit state dependent dispersive shift $\chi$ of the readout cavity determines the separation between the cavity states corresponding to
the qubit in the ground and excited state. For our sample, $\chi/2\pi = 1\:\mathrm{MHz}$ compared to the cavity linewidth $\kappa/2\pi = 4.7\:\mathrm{MHz}$, and thus each photon carries
little information about the qubit and we require a strong drive or a long integration time to distinguish the qubit state.
The measurement efficiency $\eta$ determines the number of photons observed compared to those circulating in the cavity.
Losses in the JPC amplifier, the lines connecting the readout cavity to the JPC, as well as photon losses in the cavity (both due to cavity losses and photon leakage through a second cavity port) reduce the efficiency of our measurement and thus increase the required drive strength or measurement time. The total efficiency of our measurement is $\eta=0.21$.

\section{Comparing the measured histograms with the predicted $\tau p_\textnormal{\tiny \textsc{P}}(\tau)$}
To obtain our measured histograms of $\tau p(\tau)$ we start with a one second long quantum jump trace. For each measurement that found the qubit in the ground state, we measure the length of time spanned between jumps (before and after the measurement), and increase the corresponding bin by 1. The excited state histogram is calculated similarly.

For a Poisson process with a mean time $\bar{\tau}$ between jumps, the predicted distribution is $\tau p_\textnormal{\tiny \textsc{P}}(\tau) = \frac{\tau}{\bar{\tau}}e^{-\tau / \bar{\tau}}$. 
We further normalize the distribution so that its integral is equal to the total sum of measurements in the bins. Therefore, we multiply $\tau p_\textnormal{\tiny \textsc{P}}(\tau)$ by $\Sigma\Delta/\bar{\tau}$ where $\Sigma$ is the sum of histogram values, $\bar{\tau} = \int_{-\infty}^{\infty}\tau p_\textnormal{\tiny \textsc{P}}(\tau)\,\mathrm{d}\tau$ and $\Delta$ is the size of each bin. Note that our binning was done in logarithmic scale and so the bin size depends on $\tau$, and we account for it by defining $\Delta '$ as the constant bin size in units of $\mathrm{log}(\tau)$ and adding a factor of $\mathrm{ln}(10)\tau$.
The full distribution is thus
\begin{eqnarray}
\frac{\Sigma\Delta'}{\bar{\tau}}\mathrm{ln}(10)\tau\frac{\tau}{\bar{\tau}}e^{-\tau / \bar{\tau}}. \label{eq:PoissonPrediction}
\end{eqnarray}
This is the formula used to calculate the dashed lines shown in main text Fig. 3a,b and the values $P_i$ used for $F$. Note that with this definition $\sum\limits_i{M_i}=\sum\limits_i{P_i}$ and so the fidelity $F$ is normalized in a symmetric way, and is sometimes called the Bhattacharyya coefficient\cite{Bhattacharyya1946} for probability distributions $M_i/\sum\limits_i{M_i}$ and $P_i/\sum\limits_i{P_i}$.

\section{Cross-correlation between $\bar{\tau}$ and fidelity $F$}
Main text Fig. 3c,d show measurements of $\bar{\tau}$ and $1-F$ vs. time and the correlation between these measurements signifies that low mean times between jumps correponds to non Poissonian behavior.
The correlation is more visible in supplementary Fig. 3, which shows the measured $1-F$ vs. ground state $\bar{\tau}$ for the measurements in main text Fig. 3c,d.
To quantify this visual correlation, we calculate the corresponding normalized cross-correlation:
\begin{eqnarray} 
\frac{1}{N}\sum_i\frac{(\bar{\tau}_i-\bar{\tau}_{mean})(F_i-F_{mean})}{\sigma_{\bar{\tau}}\sigma_F}, \label{eq:CrossCorr}
\end{eqnarray}
where $i$ indexes the measurements at different times shown in main text Fig. 3c,d, $\bar{\tau}_{mean}$ and $F_{mean}$ are the mean values of $\bar{\tau}_i$ and $F_i$ respectively, and $\sigma_{\bar{\tau}}$, $\sigma_F$ are their respective standard deviations. N is the total number of measurements.

We obtain a cross-correlation value of $0.87$, indicating a significant correlation. Moreover, the value of $1-F$ changes by 2 orders of magnitude between the ``quiet'' and noisy'' seconds and so the value of $F_{mean}$ is close to the ``noisy'' second values. Calculating the cross-correlation between $\bar{\tau}$ and $-\mathrm{log}(1-F)$ gives $0.98$.

 \begin{figure*} [h]

 \includegraphics[angle = 0, width = 0.5\textwidth]{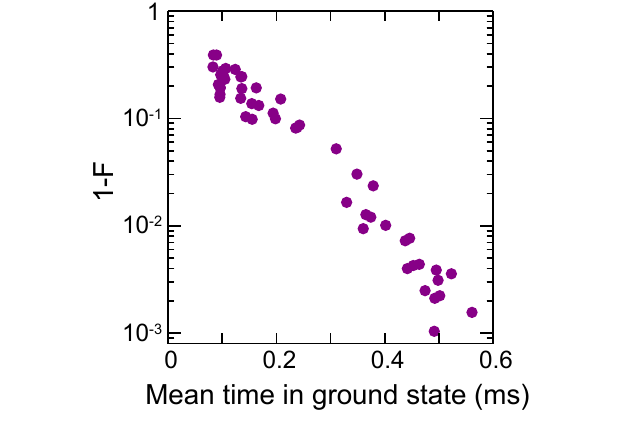}
 \caption{\label{sup_fig3}  
$1-F$ vs. ground state $\bar{\tau}$ for the measurements shown in main text Fig. 3c,d. The values are visibly anti-correlated.
 }
 \end{figure*}

\section{Spectral density of the dynamics between ``noisy'' and ``quiet'' regimes }
Tracking quantum jumps for long times, we can quantitatively analyze the statistics of the dynamics between the ``quiet'' and ``noisy'' regimes of the fluxonium qubit. In supplementary Fig. 4a we show the mean
time spent in the ground state vs. time, measured continuously for 40 hours. 
The power spectral density of this measurement is given in supplementary Fig. 4b, fitted to a power law:
\begin{eqnarray}
\frac{A}{B+(2\pi f)^{\alpha}} + C. \label{eq:SpectDens}
\end{eqnarray}
The data fits $\alpha = 1.4\pm0.1$, deviating from Poisson statistics ($\alpha = 2$) and resembling ``$1/f$'' noise.

 \begin{figure*} [h]

 \includegraphics[angle = 0, width = .7\textwidth]{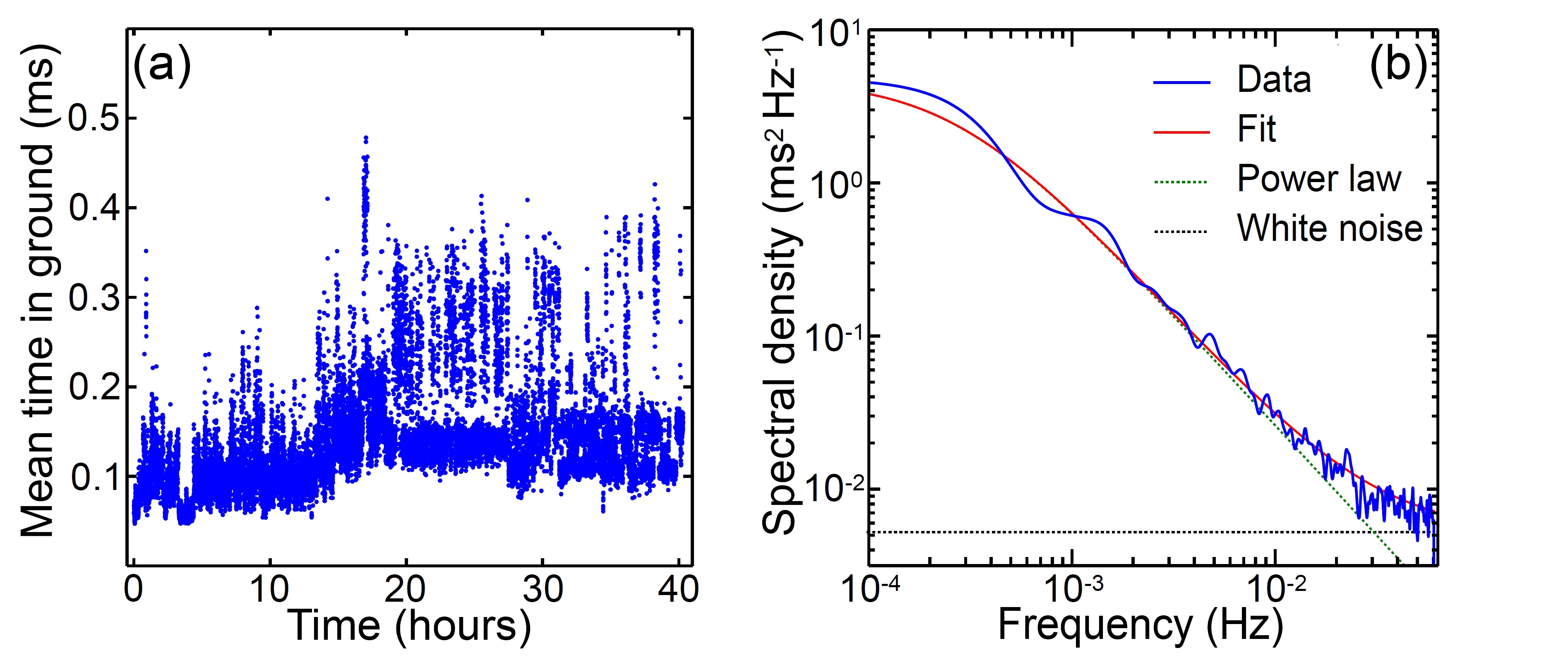}
 \caption{\label{sup_fig4}  
(a) Mean time in the ground state vs. time, measured continuously for 40 hours. Each point is an average over a 1 second quantum jump trace.
(b) Power spectral density (blue) taken over the measurement in supplementary Fig. 4a. In red is the fit to Eq. \ref{eq:SpectDens} and the green and gray dashed lines correspond to its power law and white noise components respectively.
 }
 \end{figure*}

\section{Poisson and non-Poisson behavior in the presence of  QP generation pulses and cooling in magnetic field}
In main text Fig. 4b and c we see the ground state $\bar{\tau}$ vs. time in the presence of QP generation pulses and magnetic field cooling respectively. The mean time in the ground state is  relatively short 
($<200\:\mathrm{\mu s}$) in the presence of QP generation pulses and relatively long after cooling in magnetic field ($>200\:\mathrm{\mu s}$). In supplementary Fig. 5a and b we show characteristic examples of 
$\tau p(\tau)$ distributions for the ground (blue) and excited (red) states, taken in the presence of QP generation pulses and after field cooling, with the Poisson prediction in dashed lines 
(compare to main text Fig. 3a,b). Indeed we see the correspondence between $\bar{\tau}$ and agreement with Poisson statistics. In the presence of QP generation pulses, data deviate significantly from the Poisson prediction. 

In supplementary Fig. 5c we plot the $1-F$ vs time, for measurements taken with QP generation pulses and after field cooldown, corresponding to the average times in the ground state represented in the main text 
Fig. 4b (red) and Fig. 4c (green). 
The measurements following QP generation pulses are consistently deviating from the Poisson prediction, as observed in supplementary Fig. 5a, and expected for a measurement during which the QP population is not constant. 

\begin{figure*} [h]

 \includegraphics[angle = 0, width = \textwidth]{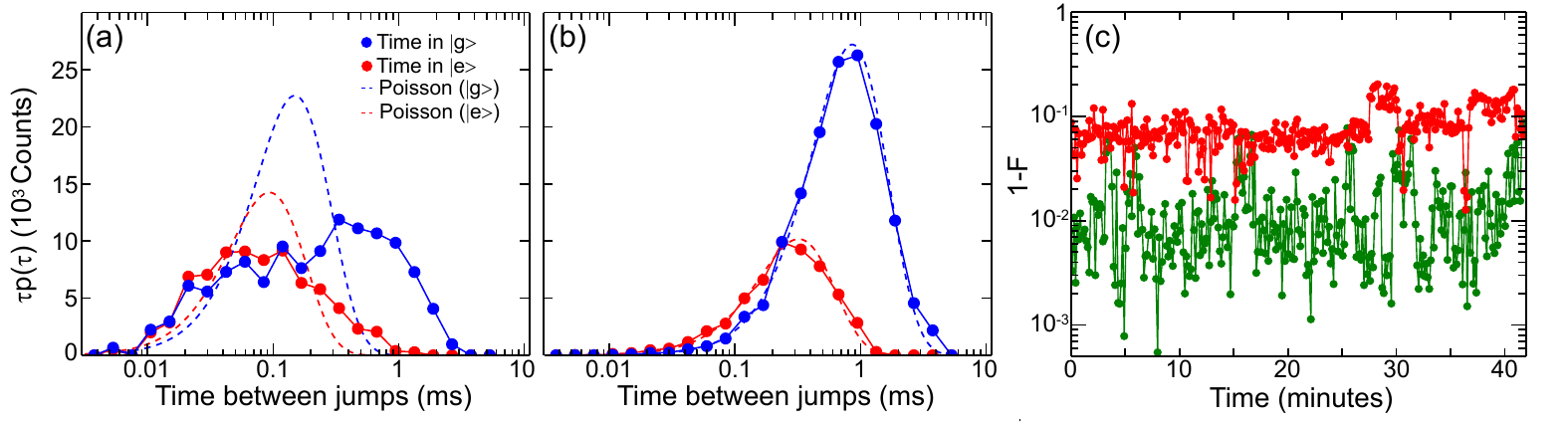}
 \caption{\label{sup_fig5}  
(a,b) Histograms with logarithmic binning of the time intervals between jumps $\tau$ for the qubit in the ground/excited (blue/red) states, scaled by $\tau$ (compare to main text Fig. 3a,b).
Data in (a) correspond to a measurement with QP generation pulses of length $t_\textnormal{\tiny \textsc{G}}=100\:\mathrm{\mu s}$ (see main text Fig. 4b) and data in (b) correspond to a measurement
after cooling in magnetic field (see main text Fig. 4c).
Data in (b) agree with the Poisson prediction, while data in (a) significantly deviate.
(c) $1-F$ vs. time in the presence of QP generation pulses (red) and field cooling (green). The corresponding ground state $\bar{\tau}$ are shown in main text Fig. 4b,c respectively. 
 }
 \end{figure*}

\section{General model for the time dependence of QP density $x_{qp}$}
Multiple mechanisms, both intrinsic and extrinsic to the superconductor, influence the time dependence of the QP density.
We can model the change in the relative QP density $x_{qp}$ using the equation\cite{Wilson2001, Wilson2004}:
\begin{eqnarray}
 \dot{x}_{qp} = g - sx_{qp} - rx_{qp}^2 \label{eq:XqpDynamics}
\end{eqnarray}
where $g$ is the QP generation coefficient, $s$ is the single QP loss due to trapping or diffusion and $r$ is the QP recombination coefficient. The parameters $g$, $s$ and $r$ have no other ambition than to describe phenomenologically an underlying microscopic physics which can be complicated. Specifically, the mechanisms causing $g$ are still hypothetical.  
We use Eq. \ref{eq:XqpDynamics} to model the equilibration of qubit lifetime after a QP generation pulse. The requilibration corresponds to a decrease of $x_{qp}$ as the injected QP leave the superinductance.
In the limit of small $x_{qp}$ deviations from the steady state value $\bar{x}_{qp}$, relevant to our experiment, the dynamics given by Eq. \ref{eq:XqpDynamics} can be linearized, and the distinction between QP mechanisms that are linear ($s$) or quadratic ($r$) in $x_{qp}$ disappears.
The value $\tau_{ss}$ shown in main text Fig. 4d then takes the value $\tau^{-1}_{ss}=s+2r\bar{x}_{qp}$.
The fits shown as a dashed line in supplementary Fig. 6a are thus based on this simpler version of the model. 
From the fits for different QP generation pulse lengths shown in the inset of main text Fig. 4d, we can extract a time constant to reach QP steady-state of $\tau_{ss} = 125\pm25\: \mathrm{ \mu s}$ and a steady-state relative QP density in the array of $\bar{x}_{qp}=4\pm1\times 10^{-8}$. This value corresponds to approximately 0.2 QP per $\mathrm{\mu m^3}$, an average population of 1-2 QP's in the whole array. From these values we can also get an average QP generation coefficient value of $g_{eff}=\bar{x}_{qp}/\tau_{ss} \approx 3\times10^{-4}\:\mathrm{s^{-1}}$.
To distinguish between the contributions of the $r$ and $s$ terms, one would need an $x_{qp}$ measurement with a larger dynamic range.
Such a measurement was recently demonstrated by $T_1$ measurements in a transmon qubit using a similar QP generation protocol\cite{Wang2014}. 

 \begin{figure*} [h]

 \includegraphics[angle = 0, width = .65\textwidth]{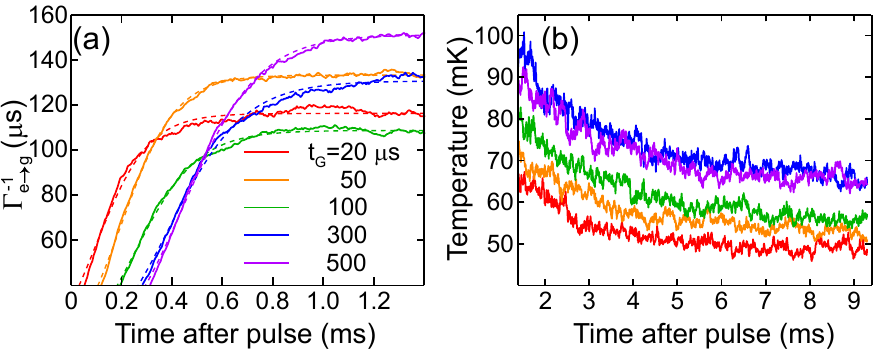}
 \caption{\label{sup_fig6}  
(a) Mean time until a jump from the excited to the ground state vs. time after QP generation pulse for different pulse lengths $t_\textnormal{\tiny \textsc{G}}$ (see legend). The dashed line is a fit to exponential ``recovery'' of qubit lifetime. This is the raw data from which main text Fig. 4d was extracted.
(b) Qubit effective temperature vs. time after QP generation pulse for pulse lengths $t_\textnormal{\tiny \textsc{G}}$ corresponding to supplementary Fig. 6a. This is the raw data from which main text Fig. 4e was extracted.
 }
 \end{figure*}
 
\section{Evaluation of temperature rise and decay following a QP generation pulse}
During the QP generation pulse, we estimate a dissipated power over the antenna junctions of $P=I_{c}*V_{2\Delta} = 10^{-10}$~$\mathrm{W}$ where $I_{c}=280$~$\mathrm{nA}$ is the critical current of the junctions and 
$V_{2\Delta}=0.4$~$\mathrm{mV}$ is the voltage corresponding to the aluminum superconducting gap. This represents a QP generation rate of  $P/2\Delta =10^6  \mathrm{\mu s^{-1}}$.
For a QP generation pulse length of $t_ \textnormal{\tiny \textsc{G}}=100\:\mathrm{\mu s}$, we estimate the total energy dissipated to be $\Delta E = P*t_\textnormal{\tiny \textsc{G}}=10^{-14}\:\mathrm{J}$.
For an estimated sapphire substrate mass of $m=0.1\:\mathrm{g}$ and an extrapolated sapphire specific heat at $50\:\mathrm{mK}$\cite{Ditmars1982} of $C=10^{-11}\:\mathrm{Jg^{-1}K^{-1}}$ the expected increase in temperature is $\Delta T = \frac{\Delta E}{Cm}\approx10\:\mathrm{mK}$. This estimate is in good agreement with the initial increase in temperature shown in the main text Fig. 4e.
Notice that the initial increase in temperature following QP generation pulse with $t_ \textnormal{\tiny \textsc{G}}=300\:\mathrm{\mu s}$ is higher than that for $t_ \textnormal{\tiny \textsc{G}}=500\:\mathrm{\mu s}$, contrary to the general trend of temperature increase with $t_ \textnormal{\tiny \textsc{G}}$. We believe this discrepancy arises from the fact that for $t_ \textnormal{\tiny \textsc{G}}>300\:\mathrm{\mu s}$ the initial temperature increase saturates and we might be observing higher order effects due to the inhomogeneity of the temperature distribution in the sapphire. In supplementary Fig. 6b we can see the raw data used, and the curves corresponding to $t_ \textnormal{\tiny \textsc{G}}=300,500\:\mathrm{\mu s}$ are separated from the others and similar to each other.

We estimate the temperature equilibration rate by $\frac{dT}{dt} = -\frac{GA}{Clm}T$ where $T$ is the temperature, $G$ is the sapphire heat conductivity, $C$ is the specific heat, $A=2.5\:\mathrm{mm^2}$ is the substrate cross section connecting to the thermal sink (the wall of the copper cavity), $l=3\:\mathrm{mm}$ is the distance to the thermal sink and $m$ is the sapphire substrate mass. Since $G$ and $C$ both have a $T^3$ temperature dependence we can take their ratio at $10\:\mathrm{K}$ (sapphire heat conductivity taken from Ref. \cite{Berman1955}).
Combining these values we get a temperature decay constant $\tau_{th}=\frac{Clm}{GA}\approx20\:\mathrm{\mu s}$. The measured temperature equilibration time of several ms shown in the main text Fig. 4e is then likely due to the sapphire-copper thermal contact.

\bibliographystyle{apsrev}
\bibliography{QPjumpsbiblio}